\documentclass[9pt,conference]{IEEEtran}
\usepackage{amssymb,amsthm,amsmath,array}
\usepackage{graphicx}
\usepackage[caption=false,font=footnotesize]{subfig}
\usepackage{xspace}
\usepackage[sort&compress, numbers]{natbib}
\usepackage{stmaryrd}
\usepackage{mathtools}
\usepackage{float}
\usepackage{textcomp}
\usepackage[table,xcdraw]{xcolor}

\begin{document}
\title{SenseAI: {\it Real-Time} Inpainting for Electron Microscopy}
\author{
\IEEEauthorblockN{
        J. Wells\IEEEauthorrefmark{1}, 
        A. Moshtaghpour\IEEEauthorrefmark{2}\IEEEauthorrefmark{3},
        D. Nicholls\IEEEauthorrefmark{2}, 
        A. W. Robinson\IEEEauthorrefmark{2}, 
        Y. Zheng\IEEEauthorrefmark{1},
        J. Castagna\IEEEauthorrefmark{3}, and
        N. D. Browning \IEEEauthorrefmark{2}\IEEEauthorrefmark{5}\IEEEauthorrefmark{6}
    }
    \IEEEauthorblockA{
        \IEEEauthorrefmark{1} Distributed Algorithms Centre for Doctoral Training, University of Liverpool, Liverpool, UK. \\
        \IEEEauthorrefmark{2} Department of Mechanical, Materials and Aerospace Engineering, University of Liverpool, UK. \\
        \IEEEauthorrefmark{3} Correlated Imaging Group, Rosalind Franklin Institute, Harwell Science and Innovation Campus, Didcot, UK. \\ 
        \IEEEauthorrefmark{4} UKRI-STFC Hartree Centre, Daresbury Laboratory, Warrington, UK. \\
        \IEEEauthorrefmark{5} Physical and Computational Science Directorate, Pacific Northwest National Laboratory, Richland, USA. \\
        \IEEEauthorrefmark{6} Sivananthan Laboratories, 590 Territorial Drive, Bolingbrook, IL, USA.
        }
        \vspace{-9mm}
}
\maketitle
\begin{abstract}
    Despite their proven success and broad applicability to Electron Microscopy (EM) data, joint dictionary-learning and sparse-coding based inpainting algorithms have so far remained impractical for real-time usage with an Electron Microscope. For many EM applications, the reconstruction time for a single "frame" is orders of magnitude longer than the data acquisition time, making it impossible to perform exclusively subsampled acquisition. This limitation has led to the development of SenseAI, a C++/CUDA library capable of extremely efficient dictionary-based inpainting. SenseAI provides $N$-dimensional dictionary learning, live reconstructions, dictionary transfer and visualization, as well as real-time plotting of statistics, parameters, and image quality metrics.
\end{abstract}

\section{Introduction}
Over the past several years, dictionary-based inpainting methods have been successfully applied to many different forms of Electron Microscopy (EM) data \cite{binev2012compressed, leary2013compressed, stevens2014potential, kovarik2016implementing, beche2016development, stevens2018subsampled, nicholls2020minimising, 
lee2021controlling, nicholls2021subsampled, robinson2022towards, nicholls2022compressive, nicholls2023targeted}. The ability of these methods to reconstruct subsampled images has a vast range of applications in the field, where sampling fewer pixels can significantly reduce the time required to perform a scan and/or reduce the electron dose incident on the sample \cite{stevens2014potential, nicholls2023targeted}. However, these algorithms are known to be computationally expensive, taking a significant amount of time to produce a single reconstruction (often on the order of minutes to hours), thus limiting their use in practical applications such as {\it real-time} reconstructions of a continuous "live" video feed directly from an electron microscope. SenseAI is a GPU-parallelised C++ library designed to overcome these limitations and support a "live" inpainting environment (i.e. visualising reconstruction results in real-time whilst inpainting a time-variable input video feed), requiring extremely efficient implementations of inpainting algorithms to provide users with frame-by-frame reconstructions, and thus allowing for the {\it real-time} operation and adjustment of a microscope in an exclusively subsampled acquisition mode.

\section{Beta-Process Factor Analysis (BPFA)}
While many dictionary-based algorithms can perform subsampled reconstructions, SenseAI employs Beta-Process Factor Analysis (BPFA) as its algorithm of choice. BPFA is a hierarchical Bayesian model that uses the two-parameter beta process (BP) \cite{ibf} to define a sparse prior on an infinite dictionary space, allowing for the non-parametric mixture modeling of latent features. For more information on BPFA and a detailed description of its model, see references \cite{nicholls2021subsampled, paisley2009nonparametric, zhou2009non}. In each iteration of the algorithm, BPFA can perform both a dictionary learning step and a sparse coding step. Using its model, it proposes an improved ("fully sampled") dictionary based solely on the subsampled input provided, making it one of the leading approaches for blind inpainting, where the dictionary and sparse weight vector are learned directly from subsampled input.

\section{N-Dimensional Inpainting}
In most typical implementations of dictionary-based inpainting algorithms for EM data, the target “image” would be 2-dimensional (2D), such as Z-contrast images from a Scanning Transmission Electron Microscope (STEM), or in some cases 3-dimensional (3D), such as RGB images or successive 2D layers of a 3D volume e.g. cryo FIB-SEM tomography \cite{nicholls2023targeted}. In these methods, for an “image” of shape $M=(M_0,M_1,M_2)$, the dictionaries used represent a set of elements with a patch shape $B$ defined only in the first two dimensions, where the size of the patch in the third dimension spans the entire shape of the target data cube ($B_2=M_2$). However, there are many examples of higher-dimensional EM data that would benefit from variable patch shapes ($B_i<<M_i$) in higher dimensions ($i > 1$). For the case of multi-frame targets such as greyscale and RGB video, this could represent training/inpainting with a single-channel dictionary to learn/reconstruct 2D patches across each layer of a multi-channel (e.g. RGB) data source, such as in the case of dictionary transfer, e.g. learning an optimal dictionary via use of STEM simulations \cite{robinson2022sim, robinson2022towards}. For hyperspectral data cubes, it may represent the use of a dictionary spanning an arbitrary n “steps” ($n<<M_2$) across a discretised spectrum. For all of the above, it may also provide a method of frame-interpolation for data cubes subsampled in the fourth (time) dimension, learning dictionary elements capturing the evolution of pixels over $f$ frames ($f << M_3$).

\begin{figure}
    \centering
    \includegraphics[width=0.47\textwidth]{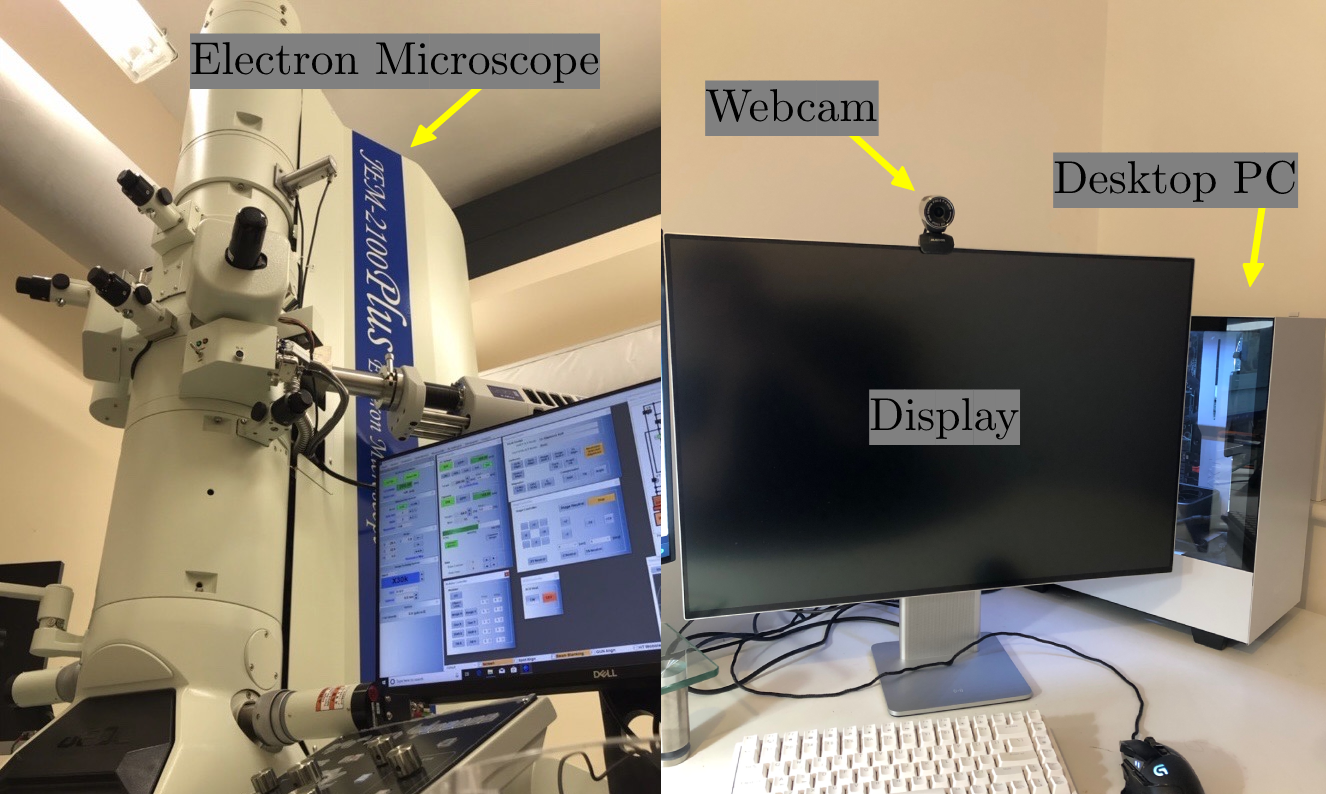}
    \caption{(left) JEOL JEM-2100 Transmission Electron Microscope. (right) The necessary set-up for a show-and-tell software DEMO}
    \label{fig:setup}
    \vspace{-4mm}
\end{figure}

\section{Time-To-Solution}
\begin{figure*}
    \centering
    \scalebox{0.18}{\includegraphics{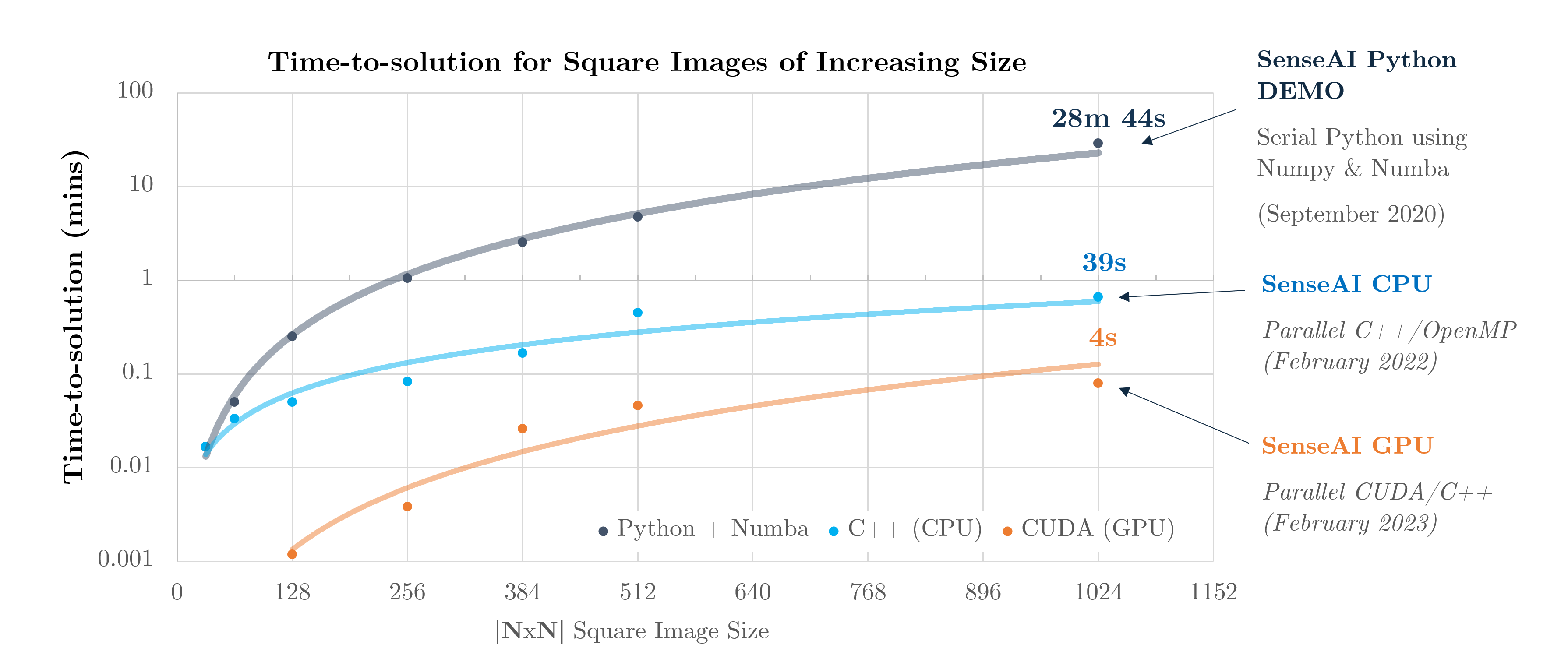}}
    \caption{Time-to-solution for images of increasing size comparing multiple different generations of SenseAI. All versions were performed on the same PC with an Intel Xeon E5-2678v3 (12 cores, 2.50 GHz), the GPU implementation was performed using an NVIDIA RTX 3060 consumer graphics card}
    \label{fig:tts}
\end{figure*}

 Fig.~\ref{fig:tts} illustrates the significant improvements in time-to-solution achieved by subsequent generations of the SenseAI code-base over the past two years. The figure shows the reconstruction times for square images of increasing size $(N,N)$, with a dictionary of 64 $(10,10)$ patches trained for two complete epochs. The total reconstruction time using each version of SenseAI was measured. As shown in Fig.~\ref{fig:tts}, the GPU-paralellised implementation has resulted in a drastic reduction in time-to-solution for common EM image sizes. For instance, for a (greyscale) $1024 \times 1024$ image, the total reconstruction time (for two complete epochs) has been reduced from approximately 29 minutes using Python/Numba to just 4 seconds.

\section{SenseAI: Feature Summary}
SenseAI has been designed to cater to the complex environment of real analytical data acquisition systems (such as in STEM), offering the following features:
\begin{itemize}
\item {\bf Modular Library Design} to easily configure varied and complicated scenarios.
\item {\bf $N$-Dimensional Inpainting} to extend beyond 2D/3D image inpainting and use arbitrarily shaped dictionary elements within BPFA algorithms.
\item {\bf Programmatic Signal Selection} for pre-determined indexing methods or on-the-fly strategic selections to improve reconstruction quality.
\item {\bf Multiple Simultaneous Problems} to perform an arbitrary number of dictionary-learning/sparse-coding problems simultaneously.
\item {\bf Dictionary Transfer between Sources} for more complex inpainting scenarios.
\item {\bf Configurable Display Window} with custom visualizations for real-time views of images, dictionaries, metrics, and more (as shown in Fig.~\ref{fig:vis}).
\end{itemize}
These features are implemented using custom CUDA kernels for maximum efficiency while remaining flexible to handle complex scenarios and use cases.

\section{Visitors Experience}
\begin{figure}[ht!]
    \centering
    \includegraphics[width=0.49\textwidth]{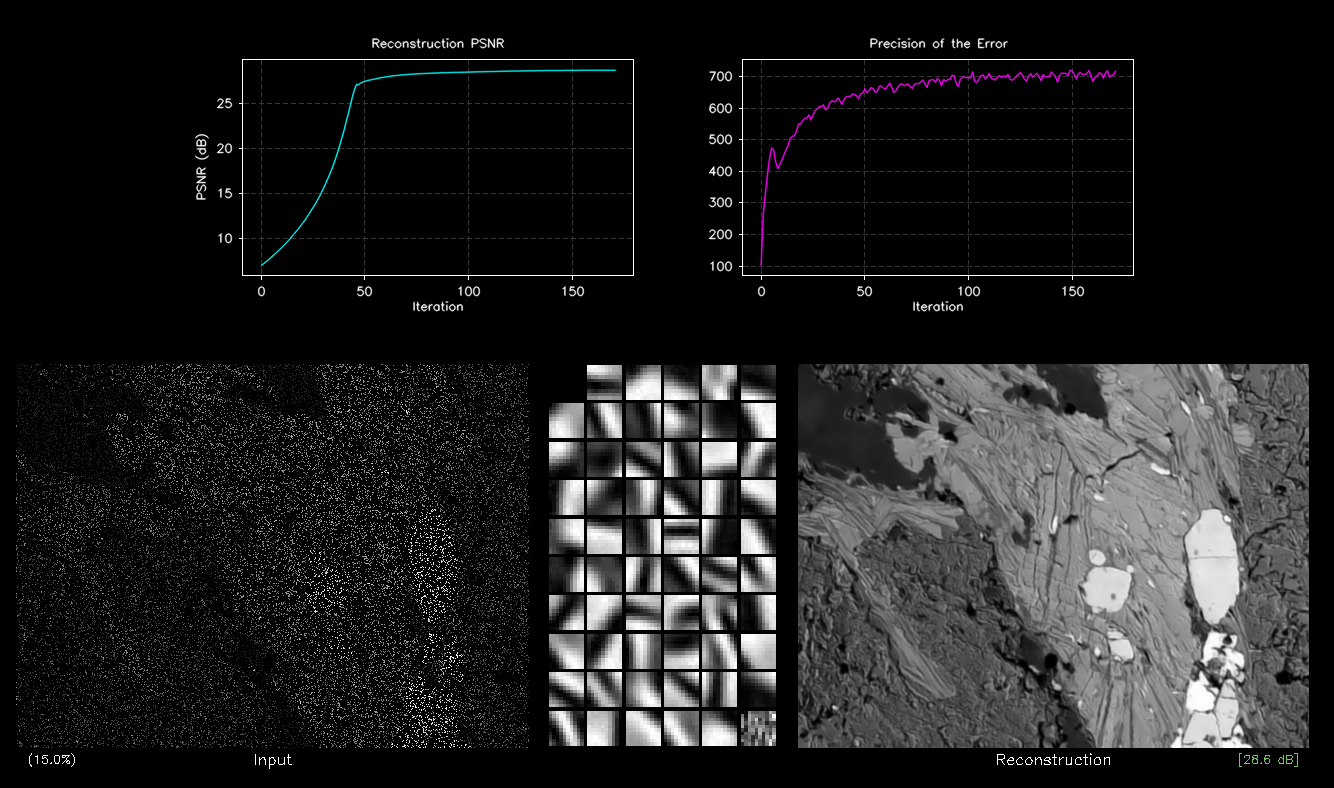}
    \caption{Example window contents for a live SEM reconstruction.}
    \label{fig:vis}
\end{figure}

SenseAI provides a configurable display window with live updates to input feeds, reconstructions, dictionaries, live plots of metrics, parameters and more, allowing for the visualisation of complex tasks as well as providing a mechanism to demonstrate the software in a practical (i.e. real-time) scenario as shown in Fig.~\ref{fig:vis}.

While SenseAI is designed to be used with a stream of data coming directly from an electron microscope, a demonstration of its capabilities can easily be performed using a desktop computer, a monitor and a standard PC webcam. Visitors will be able to observe more traditional STEM image reconstructions with real-time metrics, and a live webcam reconstruction displaying each frame of an artificially subsampled webcam stream along with reconstructed view of themselves side-by-side.

\section{Conclusion}
In this work, we introduce SenseAI, an efficient $N$-dimensional inpainting library for STEM applications, designed to overcome the limitations of using dictionary-based inpainting algorithms for real-time EM applications. In the development of SenseAI, algorithms such as BPFA have been extended into an $N$-dimensional context and optimised to be performed in parallel on a GPU, allowing for many more complex inpainting scenarios to be performed in a significantly reduced time-frame. The show-and-tell demo will showcase the software's real-time inpainting capabilities using a PC webcam, as well as more traditional STEM image reconstructions and live dictionary transfer. In the future, we aim to optimize SenseAI further for higher frame rates, showcase its higer-dimensional capabilities, such as its use in 4D-STEM, and automate parameter selection to significantly improve the ease-of-use for real-world subsampled EM applications. Overall, SenseAI promises to be a powerful tool for researchers in STEM and other fields requiring real-time, high-quality inpainting capabilities, such as the {\it real-time} operation and adjustment of an electron microscope performing exclusively subsampled acquisition.
\vfill\pagebreak

\bibliographystyle{IEEEtran}
\bibliography{refs}

\begin{thebibliography}{10}
\providecommand{\url}[1]{#1}
\csname url@samestyle\endcsname
\providecommand{\newblock}{\relax}
\providecommand{\bibinfo}[2]{#2}
\providecommand{\BIBentrySTDinterwordspacing}{\spaceskip=0pt\relax}
\providecommand{\BIBentryALTinterwordstretchfactor}{4}
\providecommand{\BIBentryALTinterwordspacing}{\spaceskip=\fontdimen2\font plus
\BIBentryALTinterwordstretchfactor\fontdimen3\font minus
  \fontdimen4\font\relax}
\providecommand{\BIBforeignlanguage}[2]{{%
\expandafter\ifx\csname l@#1\endcsname\relax
\typeout{** WARNING: IEEEtran.bst: No hyphenation pattern has been}%
\typeout{** loaded for the language `#1'. Using the pattern for}%
\typeout{** the default language instead.}%
\else
\language=\csname l@#1\endcsname
\fi
#2}}
\providecommand{\BIBdecl}{\relax}
\BIBdecl

\bibitem{binev2012compressed}
P.~Binev, W.~Dahmen, R.~DeVore, P.~Lamby, D.~Savu, and R.~Sharpley,
  ``Compressed sensing and electron microscopy,'' in \emph{Modeling Nanoscale
  Imaging in Electron Microscopy}.\hskip 1em plus 0.5em minus 0.4em\relax
  Springer, 2012, pp. 73--126.

\bibitem{leary2013compressed}
R.~Leary, Z.~Saghi, P.~A. Midgley, and D.~J. Holland, ``Compressed sensing
  electron tomography,'' \emph{Ultramicroscopy}, vol. 131, pp. 70--91, 2013.

\bibitem{stevens2014potential}
A.~Stevens, H.~Yang, L.~Carin, I.~Arslan, and N.~D. Browning, ``The potential
  for bayesian compressive sensing to significantly reduce electron dose in
  high-resolution stem images,'' \emph{Microscopy}, vol.~63, no.~1, pp. 41--51,
  2014.

\bibitem{kovarik2016implementing}
L.~Kovarik, A.~Stevens, A.~Liyu, and N.~D. Browning, ``Implementing an accurate
  and rapid sparse sampling approach for low-dose atomic resolution stem
  imaging,'' \emph{Applied Physics Letters}, vol. 109, no.~16, p. 164102, 2016.

\bibitem{beche2016development}
A.~B{\'e}ch{\'e}, B.~Goris, B.~Freitag, and J.~Verbeeck, ``Development of a
  fast electromagnetic beam blanker for compressed sensing in scanning
  transmission electron microscopy,'' \emph{Applied Physics Letters}, vol. 108,
  no.~9, p. 093103, 2016.

\bibitem{stevens2018subsampled}
A.~Stevens, H.~Yang, W.~Hao, L.~Jones, C.~Ophus, P.~D. Nellist, and N.~D.
  Browning, ``Subsampled stem-ptychography,'' \emph{Applied Physics Letters},
  vol. 113, no.~3, p. 033104, 2018.

\bibitem{nicholls2020minimising}
D.~Nicholls, J.~Lee, H.~Amari, A.~J. Stevens, B.~L. Mehdi, and N.~D. Browning,
  ``Minimising damage in high resolution scanning transmission electron
  microscope images of nanoscale structures and processes,'' \emph{Nanoscale},
  vol.~12, no.~41, pp. 21\,248--21\,254, 2020.

\bibitem{lee2021controlling}
J.~Lee, D.~Nicholls, N.~D. Browning, and B.~L. Mehdi, ``Controlling radiolysis
  chemistry on the nanoscale in liquid cell scanning transmission electron
  microscopy,'' \emph{Physical Chemistry Chemical Physics}, vol.~23, no.~33,
  pp. 17\,766--17\,773, 2021.

\bibitem{nicholls2021subsampled}
D.~Nicholls, J.~Wells, A.~Stevens, Y.~Zheng, J.~Castagna, and N.~D. Browning,
  ``Sub-sampled imaging for stem: Maximising image speed, resolution and
  precision through reconstruction parameter refinement,'' \emph{Submitted},
  2021.

\bibitem{robinson2022towards}
A.~Robinson, J.~Wells, D.~Nicholls, A.~Moshtaghpour, M.~Chi, A.~Kirkland, and
  N.~Browning, ``Towards real-time stem simulations through targeted
  sub-sampling strategies,'' \emph{Journal of microscopy}, 02 2023.

\bibitem{nicholls2022compressive}
D.~Nicholls, A.~Robinson, J.~Wells, A.~Moshtaghpour, M.~Bahri, A.~Kirkland, and
  N.~Browning, ``Compressive scanning transmission electron microscopy,'' in
  \emph{ICASSP 2022 - 2022 IEEE International Conference on Acoustics, Speech
  and Signal Processing (ICASSP)}, 2022, pp. 1586--1590.

\bibitem{nicholls2023targeted}
D.~Nicholls, J.~Wells, A.~Robinson, A.~Moshtaghpour, M.~Kobylynska, R.~Fleck,
  A.~Kirkland, and N.~Browning, ``A targeted sampling strategy for compressive
  cryo focused ion beam scanning electron microscopy,'' in \emph{ICASSP 2023 -
  2023 IEEE International Conference on Acoustics, Speech and Signal Processing
  (ICASSP)}, 11 2022.

\bibitem{ibf}
\BIBentryALTinterwordspacing
R.~Thibaux and M.~I. Jordan, ``Hierarchical beta processes and the indian
  buffet process,'' in \emph{Proceedings of the Eleventh International
  Conference on Artificial Intelligence and Statistics}, ser. Proceedings of
  Machine Learning Research, M.~Meila and X.~Shen, Eds., vol.~2.\hskip 1em plus
  0.5em minus 0.4em\relax San Juan, Puerto Rico: PMLR, 21--24 Mar 2007, pp.
  564--571. [Online]. Available:
  \url{https://proceedings.mlr.press/v2/thibaux07a.html}
\BIBentrySTDinterwordspacing

\bibitem{paisley2009nonparametric}
J.~Paisley and L.~Carin, ``Nonparametric factor analysis with beta process
  priors,'' in \emph{Proceedings of the 26th annual international conference on
  machine learning}, 2009, pp. 777--784.

\bibitem{zhou2009non}
M.~Zhou, H.~Chen, J.~W. Paisley, L.~Ren, G.~Sapiro, and L.~Carin,
  ``Non-parametric {B}ayesian dictionary learning for sparse image
  representations.'' in \emph{NIPS}, vol.~9, 2009, pp. 2295--2303.

\bibitem{robinson2022sim}
A.~Robinson, D.~Nicholls, J.~Wells, A.~Moshtaghpour, A.~Kirkland, and N.~D.
  Browning, ``{SIM-STEM} {L}ab: Incorporating compressed sensing theory for
  fast {STEM} simulation,'' \emph{Ultramicroscopy}, p. 113625, 2022.

\end{thebibliography}
\end{document}